\documentclass[preprint,superscriptaddress,showkeys,showpacs,nofootinbib,byrevtex,amsmath,showpreprintmunbers]{revtex4} 
\usepackage{graphicx}
\usepackage{bm}
\begin{document}      
\preprint{YITP-08-72}
\preprint{INHA-NTG-15/08}
\title{Pion properties at finite density}
\author{Seung-il~Nam}
\email[E-mail: ]{sinam@yukawa.kyoto-u.ac.jp}
\affiliation{Yukawa Institute for Theoretical Physics, Kyoto University,\\
Kitashirakawaoiwake, Sakyo, Kyoto 606-8502, Japan} 
\author{Hyun-Chul~Kim}
\email[E-mail: ]{hchkim@inha.ac.kr}
\affiliation{Department of Physics, Inha University,\\ 
253 Yonghuyn-dong, Nam-gu Incheon 402-751, Korea} 
\date{\today}
\begin{abstract} 
In this talk, we report our recent work on the pion weak decay
constant ($F_\pi$) and pion mass ($m_\pi$) using the nonlocal 
chiral quark model with the finite quark-number 
chemical potential ($\mu$) taken into account. Considering the
breakdown of Lorentz invariance at finite density, the time and space
components are computed separately, and the corresponding 
results turn out to be: $F^t_\pi=82.96$ MeV and $F^s_\pi=80.29$ MeV at
$\mu_c\approx320$ MeV, respectively. Using the in-medium
Gell-Mann-Oakes-Renner (GOR) relation, we show that the pion mass
increases by about $15\%$ 
at $\mu_c$.   
\end{abstract} 
\maketitle
\section{Introduction}	
The pion is identified as the Nambu-Goldstone (NG) boson from the
spontaneous breakdown of chiral symmetry (SB$\chi$S) which is 
essential in describing low-energy hadronic phenomena. Thus, the
in-medium modification of the $F_\pi$ and $m_\pi$ is of great
importance to understand the chiral symmetry restoration in
matter. Experimentally, the modifications of these quantities can be
measured from deeply bound pionic atoms. 

There has been a great amount of theoretical works on this
subject. For example, meson-baryon chiral perturbation theory
($\chi$PT) and models with chiral  
symmetry were applied for this
purpose~\cite{Kirchbach:1993ep,Wirzba:1995sh,Kim:2003tp}. Since the
Lorentz invariance is broken in 
medium, one has to study the space and time components of the pion
weak decay constant separately.  In in-medium $\chi$PT~\cite{Kirchbach:1997rk},   
the magnitude of its space component $F^s_\pi$ was shown to be about 
four times smaller than that of the time component $F^t_\pi$ at
$\rho_0$. In the QCD sum rules, it was discussed that dimension-five
operators are responsible for making splitting between $F^t_\pi$ and
$F^s_\pi$, and the contributions of the intermediate $\Delta$ state
makes $F^s_\pi$ much smaller than $F^t_\pi$~\cite{Kim:2003tp}.

In the present work, we investigate the modifications of
the $F_\pi$ and $m_\pi$ for the finite quark-number chemical potential
($\mu\ne0$) but at zero temperature ($T=0$), 
employing the nonlocal chiral quark model (NL$\chi$QM), derived
from the nontrivial instanton
vacuum~\cite{Diakonov:1985eg,Carter:1998ji}.  We will show that
$F^t_\pi=82.96$ MeV 
and $F^s_\pi=80.29$ MeV at the critical value $\mu_c\approx320$ MeV,
which are about $13\sim16\%$ smaller than that in free space
($F_\pi=93$ MeV).  The results are 
compatible with those obtained in other models.  Using the GOR
relation, satisfied within the model~\cite{Diakonov:1985eg}, we
estimate the pion mass shift and find that the mass is increased by
about $15\,\%$ at $\mu_c$.    

\section{Nonlocal chiral quark model with finite $\mu$}	
The quark zero-mode solution in the presence of the instanton can be
obtained with the finite $\mu$ as follows~\cite{Carter:1998ji}:  
\begin{equation}
\label{eq:ZM}
\left[i\rlap{/}{\partial}-i\gamma_4\mu
-\rlap{/}{A}_{I\bar{I}}\right]\Psi^{(0)}_{I\bar{I}}
=0.
\end{equation}
A quark propagator deduced from Eq.~(\ref{eq:ZM}) reads: 
\begin{equation}
\label{eq:prop2}
S=\frac{1}{i\rlap{/}{\partial}-i\gamma_4\mu+iM(i\partial,\mu)},
\end{equation}
where $M$ denotes the momentum-dependent and $\mu$-dependent quark
mass that arises from the Fourier transform of the quark zero-mode
solution:  
\begin{equation}
\label{eq:MDQM}
M(\bar{k})=M_0(\mu)\bar{k}^2\psi^2(\bar{k}).  
\end{equation}
Here, $\bar{k}=(\vec{k},k_4+i\mu)$.  The $M_0$ is the constituent
quark mass at $k^2=0$, which depends on $\mu$ and becomes zero at the
critical value $\mu_c\approx320$ MeV, indicating the first-order phase
transition~\cite{Nam:2008xx}. The analytical expressions for 
$\psi_4$ and $\vec{\psi}$ are given in Ref.~6.  Since the Lorentz
invariance is broken for the finite $\mu$, the PCAC relation in medium
should be decomposed into the space ($s$) and time ($t$) parts 
as follows: 
\begin{equation}
\label{eq:PCACst}
\langle0|{\bm A}^a(x)|\pi^b(P)\rangle=i\sqrt{2}F^s_\pi
\delta^{ab} {\bm P}e^{-iP\cdot x},\,\,
\langle0|A^a_4(x)|\pi^b(P)\rangle=i\sqrt{2}F^t_\pi\delta^{ab}
P_4e^{-iP\cdot x}.
\end{equation}
In order to compute the relevant matrix elements, we write the
effective chiral action in the presence of an external axial-vector
source $J^a_{5\mu}$:  
\begin{eqnarray}
\label{eq:ECA}
\mathcal{S}_\mathrm{eff}[\pi,\mu,J^a_{5\mu}]&=&
-\mathrm{Sp}\ln
\left[i\rlap{/}{\bar{\partial}}
+\gamma_5\gamma^{\mu}\frac{\tau^a}{2}J^a_{5\mu}
+\sqrt{M(i\bar{\partial},J^a_{5\mu})} U_5
\sqrt{M(i\bar{\partial},J^a_{5\mu})} \right]. 
\end{eqnarray}
From this effective chiral action and the in-medium PCAC relation, we can obtain the following compact expressions for the $F_\pi$ for finite $\mu$:
\begin{eqnarray}
\label{eq:}
F^s_\pi(\mu)&\approx&F^\mathrm{exp}_\pi+\mu^2\left[\frac{N_c}
{F^\mathrm{exp}_\pi}
\int\frac{d^4k}{(2\pi)^4}\left( 
\frac{8k^2_4\tilde{\mathcal{M}}'\tilde{\mathcal{M}}''}{k^2+\mathcal{M}^2}
-\frac{10k^2_4\tilde{\mathcal{M}}'^2}{[k^2+\mathcal{M}^2]^2}
\right)\right],
\cr
F^t_\pi(\mu)&\approx&F^s_\pi(\mu)+\mu^2\left[\frac{N_c}
{F^\mathrm{exp}_\pi}
\int\frac{d^4k}{(2\pi)^4}
\frac{8k^2_4\tilde{\mathcal{M}}'^2}{[k^2+\mathcal{M}^2]^2}\right].
\end{eqnarray}
For  more details, one can refer Ref.~7.
\section{Numerical results}	
\begin{figure}[t]
\begin{tabular}{cc}
\includegraphics[width=7cm]{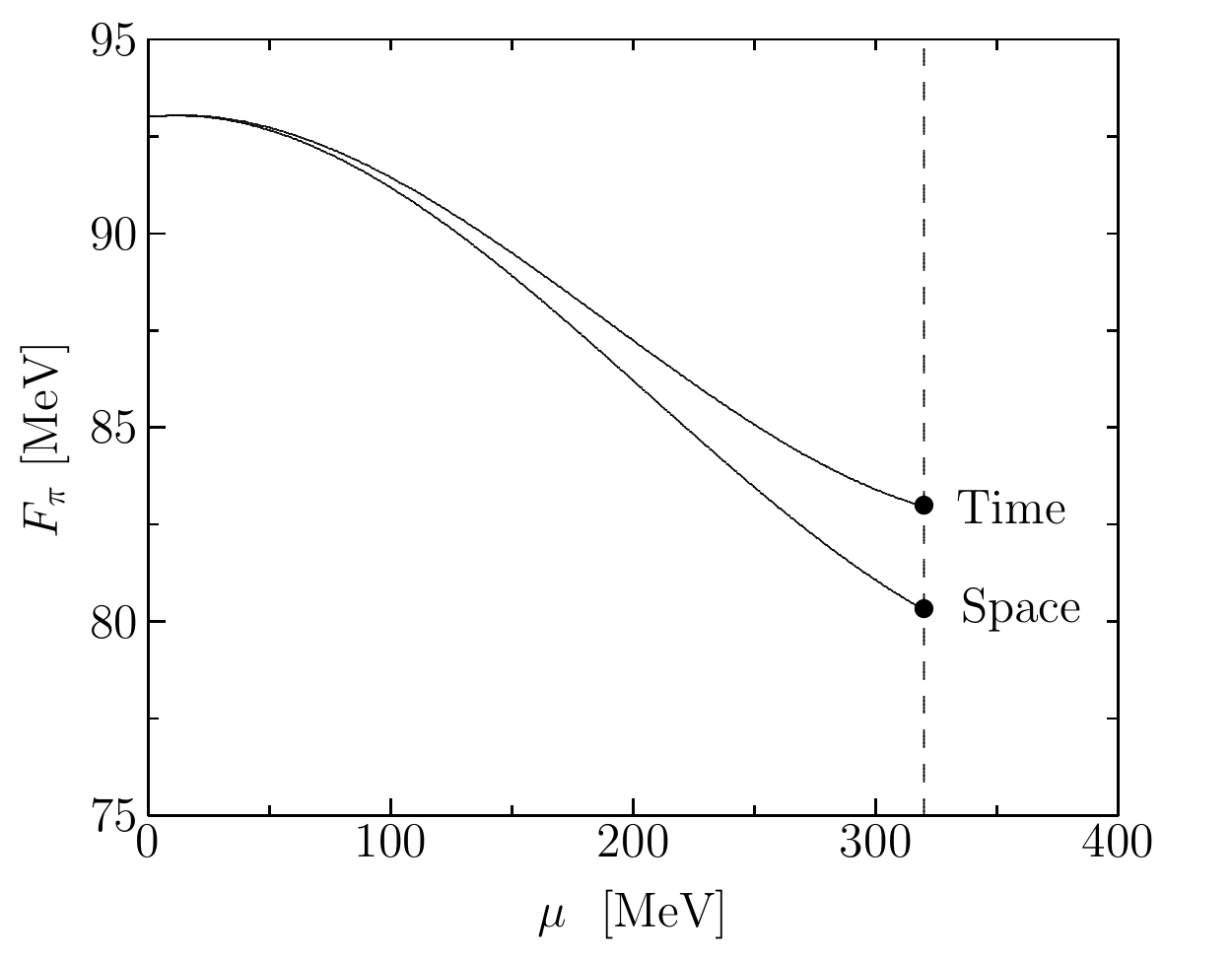}
\includegraphics[width=7cm]{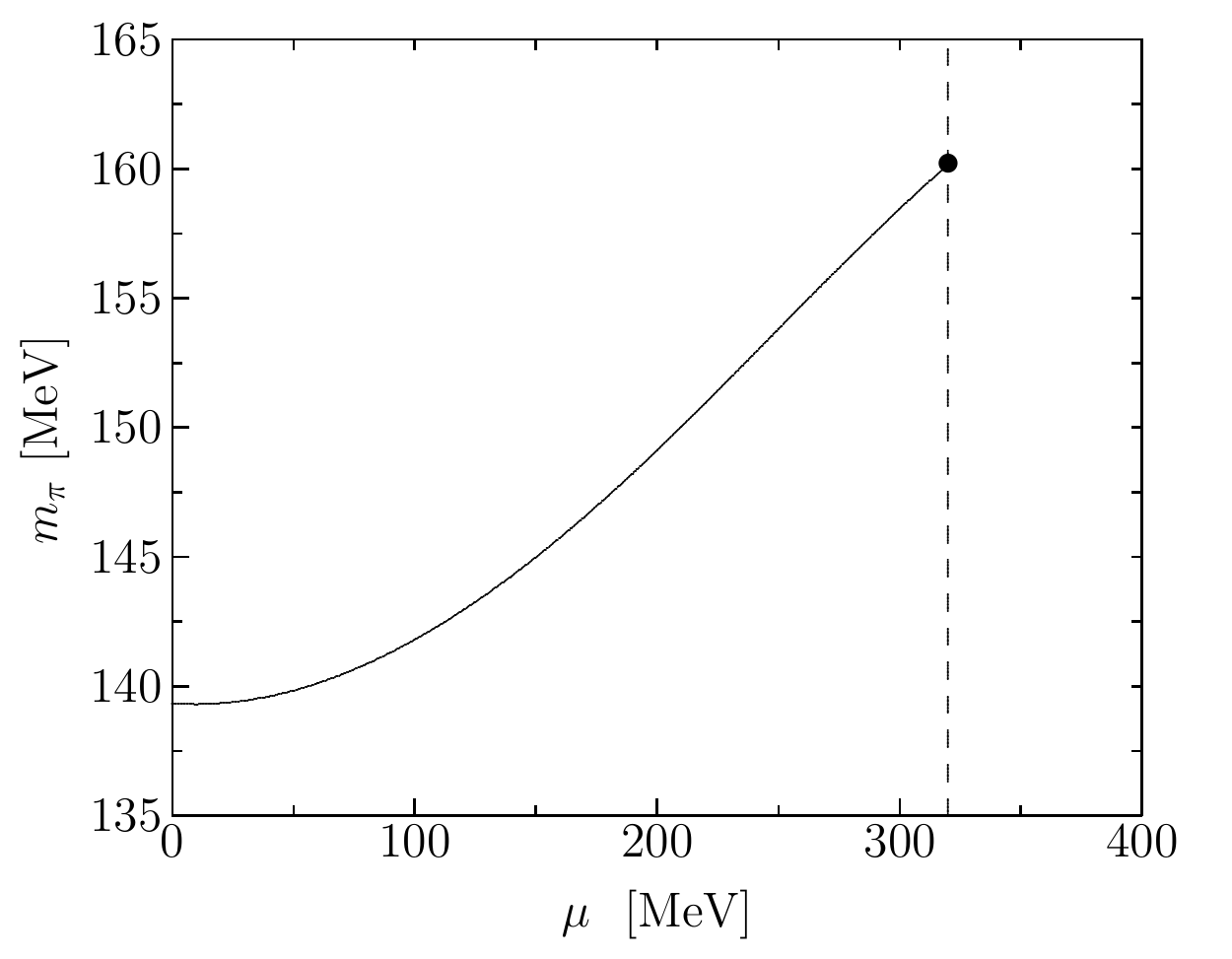}
\end{tabular}
\caption{$F^s_\pi$ and $F^t_\pi$ (left), and $m_\pi$ (right) as
  functions of the quark-number chemical potential $\mu$ up to $\mu=\mu_c\approx320$ MeV.}       
\label{fig1}
\end{figure}

We now present numerical results for the $F_\pi$ for finite $\mu$. In the left panel of Fig.~\ref{fig1}, we
show $F^s_\pi$ and $F^t_\pi$ as functions of $\mu$. At $\mu=0$ we find $F^s_\pi=F^t_\pi\approx93$ MeV as it should
be, and we obtain $F^t_\pi\approx 82.96$ MeV and $F^s_\pi\approx
80.29$ MeV at $\mu_c\approx 320$ MeV. When we examine the ratio
$F^{(s,t)}/F^\mathrm{exp}_\pi$, it must be unity at $\mu=0$ and then
it is getting smaller gradually as $\mu$ increases.  At $\mu_c$, it turns out that $F^t_\pi/F^\mathrm{exp}_\pi
\approx 0.89$ and $F^s_\pi / F^\mathrm{exp}_\pi \approx 0.86$. From
these observations, $F^s_\pi/F^t_\pi$ is less than unity
for the whole region of the NG phase ($\mu\le\mu_c$), and the $F_\pi$ is reduced by
about $13\sim16\%$. We summarize the results in Table~\ref{TB}. In the QCD sum
rule~\cite{Kim:2003tp}, it was discussed that the splitting between 
the time and space components is represented by the dimension-five
condensates.  Especially, $F^s_\pi/F^\mathrm{exp}_\pi$
becomes much smaller, when the intermediate $\Delta$ state is
considered ($0.78\to0.57$).  The ratios were also studied in in-medium
$\chi$PT in the heavy baryon limit~\cite{Kirchbach:1997rk}: It was 
found that $F^t_\pi/F^\mathrm{exp}_\pi\approx0.90$, which is
compatible with the present results, whereas the space component is
given much smaller: $F^s_\pi/F^\mathrm{exp}_\pi\approx0.25$ at
$\rho_0$.  

\begin{table}[h]
\caption{Numerical results for $F^{s,t}_\pi$ and $m_\pi$ for finite $\mu$.}
\begin{tabular}{c|c|c|c} 
&$F^s_\pi$&$F^t_\pi$&$m_\pi$\\
\hline
$\mu=0$&$93$ MeV&$93$ MeV&$139.33$ MeV\\
\hline
$\mu_c\approx320$ MeV&$80.29$ MeV&$82.96$ MeV&$160.14$ MeV\\
\hline
Modification&$16\%\downarrow$&$13\%\downarrow$&$15\%\uparrow$\\
\end{tabular}
\label{TB}
\end{table}

Now, we would like to investigate the change of the pion mass for finite $\mu$.  For  this purpose, we use the results for $F^{s,t}_\pi$ computed previously and the in-medium GOR relation:  
\begin{equation}
\label{eq:GOR}
(m^*_\pi F^t_\pi)^2=2m_q\langle{iq^{\dagger}q}\rangle^*,
\end{equation}
where $m_q$ is the current quark mass, taken to be around $5$
MeV.  In the right panel of
Fig.~\ref{fig1}, we draw $m_\pi$ as a function of $\mu$. In free
space, we observe $m_\pi=139.33$ MeV which is in good agreement 
with the experimental value, whereas $m_\pi=160.14$ MeV at the
critical value $\mu_c\approx 320$ MeV.  This observation tells us that
the pion mass increases almost linearly and  at $\mu_c$ it becomes
about $15\%$ heavier than that in free space.  We summarize this
result also in Table~\ref{TB}.  
\section{Summary and conclusion}	
We have investigated the pion weak decay constant and pion mass for
finite $\mu$ within the framework of the nonlocal chiral quark model
in the presence of the finite quark-number chemical potential.  For
the numerical results, we obtained $F^t_\pi=82.96$ MeV and
$F^s_\pi=80.29$ MeV at $\mu_c\approx320$ MeV. Considering the
in-medium Gell-Mann-Oakes-Renner relation, we also studied the pion
mass modification for finite $\mu$.  The results are summarized in
Table~\ref{TB}.   However, the splitting between the time and space 
components of $F_\pi$ has turned out to be relatively small in
comparison to those from in-medium chiral perturbation theory and the
QCD sum rule.  As discussed in the previous section, the
$\Delta$-state contribution  may cause this difference. A further
study including meson-loop corrections is 
under progress.

\section*{Acknowledgments}
The authors are grateful to the organizers for the 4th Asia-Pacific
Conference on Few-Body Problems in Physics 2008 (APFB08), which was
held during $19\sim23$ August 2008, in Depok, Indonesia. They also
thank T.~Kunihiro and S.~H. Lee for fruitful discussions. S.i.N. was
partially supported by the Grant for Scientific Research (Priority
Area No.17070002 and No.20028005) from the Ministry of Education,
Culture, Science and Technology (MEXT) of
Japan.  The present work is also supported by Inha University Research
Grant (INHA-37453). 


\end{document}